\documentclass[3p,times,twocolumn]{elsarticle}

\usepackage{ecrc}
\usepackage{multicol,caption}

\volume{00}

\firstpage{1}

\journalname{Nuclear and Particle Physics Proceedings}

\runauth{}

\jid{nppp}

\jnltitlelogo{Nuclear and Particle Physics Proceedings}

\usepackage{amssymb}

\usepackage[figuresright]{rotating}

\begin{document}

\begin{frontmatter}



\fntext[label1]{On behalf of the NA62 Collaboration:
G.~Aglieri Rinella, R.~Aliberti, F.~Ambrosino, R.~Ammendola, B.~Angelucci, A.~Antonelli, 
G.~Anzivino, R.~Arcidiacono, I.~Azhinenko, 
S.~Balev
% ...
}


\dochead{}

\title{Recent results and prospects for NA62 experiment}


\author{Silvia Martellotti\fnref{label1}}

\address{INFN, Laboratory Nazionali di Frascati (RM), Italy}
\ead{silvia.martellotti@lnf.infn.it}

\begin{abstract}
The $K^+\rightarrow \pi^+ \nu \bar \nu$ decay is theoretically one of the cleanest meson 
decays and so a good place to look for indirect effects of new physics complementary to LHC searches.
The NA62 experiment at CERN is designed to measure the branching ratio of this decay with 10\% precision. NA62 was commissioned in October 2014, took data in pilot runs in 2014 and 2015. The NA62 experimental setup is illustrated and data quality is reported.
\end{abstract}


\end{frontmatter}


\section{The NA62 Experiment}
\label{}

The NA62 (62nd experiment in the CERN North Area) \cite{NA62} is a fixed target experiment at CERN operating in the 400GeV/c proton beam supplied by the CERN Super-Proton-Synchrotron (SPS) facility. The goal of the experiment is to test the Standard Model (SM). A secondary $K^+$ beam with momentum $p_{K^+} = 75$ GeV/c is selected. The NA62 sub-detectors are located along the trajectory of the $K^+$ beam. They aim to identify kaon decay particles and to measure their momentum and energy. The NA62 experiment had a pilot run in October-November 2014. Data collected in the pilot run were used to study the detector performance and to validate the analysis method. The NA62 experiment plans to collect data each year until 2018, when the second long shutdown of the Large Hadron Collider (LHC) machine is foreseen.

\section{$BR(K^+\rightarrow \pi^+ \nu \bar \nu$) measurement}

\begin{figure}[h]
\includegraphics[width=7.8cm]{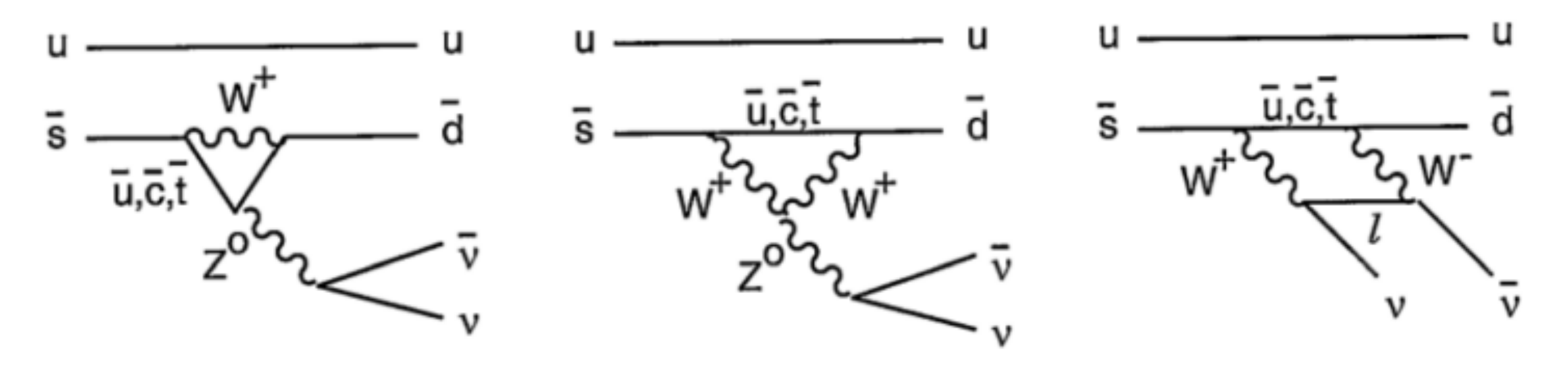}
\caption{Box diagram and Z-penguin diagrams contributing to the process $K\rightarrow \pi\nu\bar\nu$.}\label{fig:1}
\end{figure}

The $K^+\rightarrow \pi^+ \nu \bar \nu$ and $K_L\rightarrow \pi^0 \nu \bar \nu$ are 
flavour changing neutral current decays proceeding through box and electroweak penguin diagrams. A quadratic GIM mechanism and strong Cabibbo suppression make these processes extremely rare. Using the value of tree-level elements of the Cabibbo-Kobayashi-Maskawa (CKM) triangle as external inputs, the Standard Model (SM) predicts \cite{ref:carico}\cite{ref:neutro}:
\begin{displaymath}
BR(K^+\rightarrow \pi^+ \nu \bar \nu) = (8.4 \pm 1.0) \times 10^{-11}
\end{displaymath}
\begin{displaymath}
BR(K_L\rightarrow \pi^0 \nu \bar \nu) = (3.4 \pm 0.6) \times 10^{-11}
\end{displaymath}

The theoretical accuracy is at the percent level, because short distance physics
dominates thanks to the top quark exchange in the loop. The hadronic matrix elements cancel
almost completely in the normalization of the $K \rightarrow \pi\nu\bar\nu$
branching ratios to the precisely measured BR of the $K^+ \rightarrow \pi^0 e^+ \nu_e$.
Experimental knowledge of the external inputs dominate the uncertainties on these predictions. 
The dependence on CKM parameters partially cancels in the correlation between 
$K^+\rightarrow \pi^+ \nu \bar \nu$ and $K_L\rightarrow \pi^0 \nu \bar \nu$.
Therefore simultaneous measurements of the two BRs' would allow a theoretical clean investigation of the CKM triangle using kaons only. The $K \rightarrow \pi\nu\bar\nu$ decays are
extremely sensitive to physics beyond the SM, probing the highest mass scales among the rare 
meson decays. The largest deviations from SM are expected in
models with new sources of flavour violation, owing to weaker constraints from B physics 
\cite{ref:teoria1}\cite{ref:teoria2}.
The experimental value of $\epsilon_K$
the parameter measuring the indirect CP violation in neutral kaon decays, limits the range of variation expected for $K \rightarrow \pi\nu\bar\nu$ BRs within 
models with currents of defined chirality, producing typical correlation patterns between charged and neutral
modes\cite{ref:teoria3}. 
Results from LHC direct searches strongly limit the range of variation, mainly in supersymmetric models 
\cite{ref:ss1}\cite{ref:ss2}. 
Anyway, thanks to SM suppression and existing constraints from K physics, signifcant variations of the $K \rightarrow \pi\nu\bar\nu$ BRs from the SM predictions
induced by new physics at mass scales up to 100 TeV are still observable by experiment with at least 10\% precision.

The most precise experimental result has been obtained by the dedicated experiments E787 and E949 at the Brookhaven National Laboratory\cite{ref:misura1}\cite{ref:misura2} which collected a total of 7 events using a decay-at-rest technique. Only the charged mode has been observed so far, and the present status is
\cite{ref:misura3}:

\begin{displaymath}
BR(K^+\rightarrow \pi^+ \nu\bar \nu) = (17.3^{+11.5}_{-10.5}) \times 10^{-11}
\end{displaymath}
\begin{displaymath}
BR(K^+\rightarrow \pi^+ \nu\bar \nu)  < 2.6 \times 10^{-8}  \; 90\% CL.
\end{displaymath}
still far from the precision of the SM prediction.

\section{The NA62 apparatus}
\label{}

\begin{figure}[h]
\centering
\includegraphics[width=8.2cm]{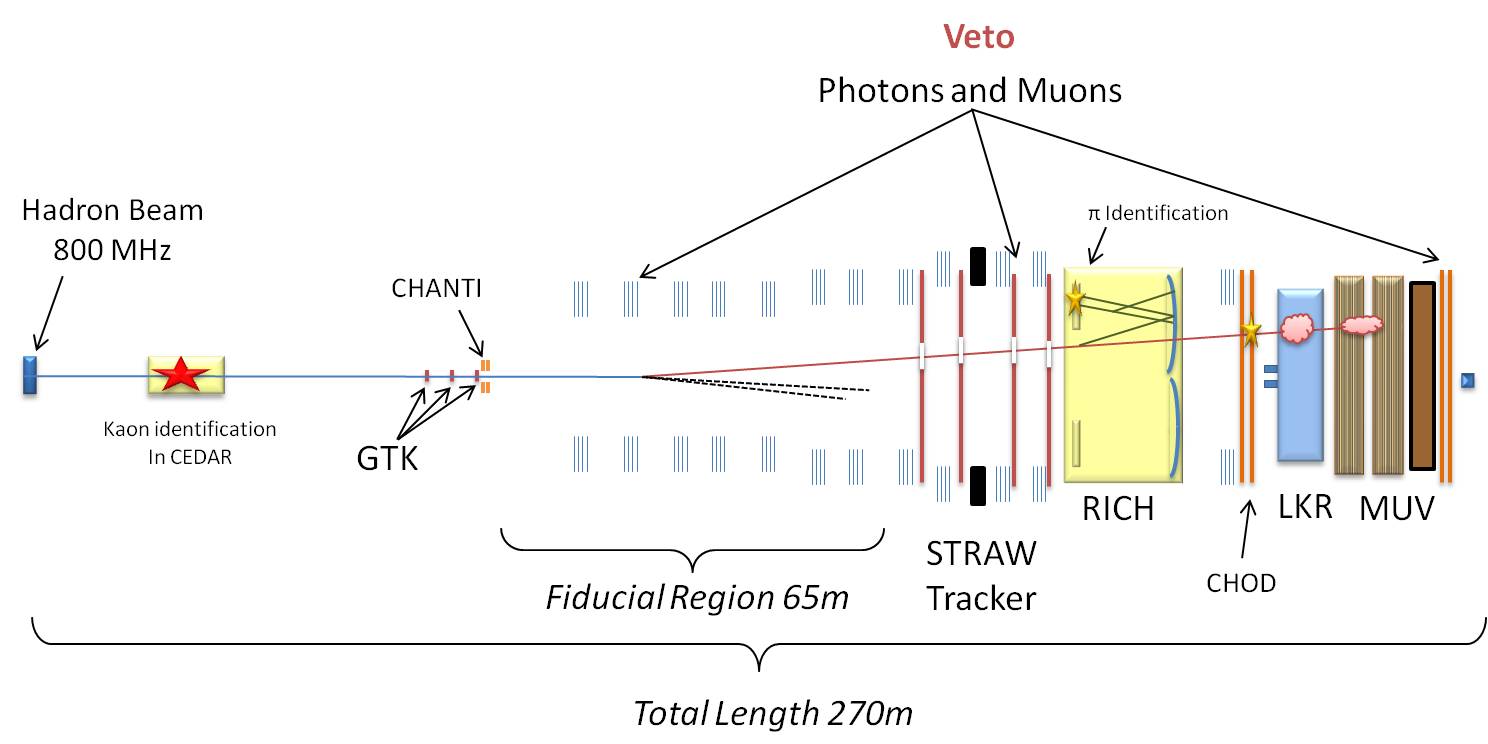}
\caption{Scheme of the NA62 layout.}
\label{fig:setup}
\end{figure}

A measurement of the BR of the $BR(K^+\rightarrow \pi^+ \nu\bar \nu)$ with 10\% precision is the main goal
of the NA62 experiment at CERN\cite{ref:exp1}\cite{ref:exp2}. The experiment plans to collect about $10^{13}$ kaon decays in a few years using 400 GeV/c protons from SPS. The design requirement then are  10\% signal acceptance and of
the order of 10\% signal to background ratio\cite{ref:design}. The proton energy and experimental requirement force the use of a kaon decay in flight technique. The 12 orders of magnitude of background rejection requires the use of as 
much as possible independent experimental techniques to suppress unwanted final states.

Figure \ref{fig:setup} shows the NA62 apparatus for $K^+\rightarrow \pi^+ \nu\bar\nu$ search. 
Protons impinge on a Be target producing secondary charged particles, 6\% of which are kaons. A 100 m long beam line selects, collimates, focuses and transports charged particles of $75 \pm 0.8$ GeV/c momentum, down to a 100 m long evacuated decay region. A Cerenkov counter (KTAG) filled with $N_2$ along the beam line identifes and timestamps kaons. Three silicon pixel stations
(Gigatracker) of $6 \times 3$ cm$^2$ surface trace and timestamp all the beam particles before they enter
the vacuum region downstream. The Gigatracker faces the full 750 MHz beam rate, the
KTAG the rate from kaons only. A guard ring detector (CHANTI) tags hadronic interactions
at the entrance of the decay volume. About 10\% of kaons decay in the vacuum region between
the GTK and the downstream detectors. Large Angle annular electromagnetic calorimeters
(LAV) made of lead glass blocks surround the decay and downstream volumes to catch off-axis photons
out to 50 mrad. A magnetic spectrometer of straw tubes in vacuum traces charged particles. Holes of
variable radius from 6.5 to 13 cm around the beam axis in the spectrometer chambers and in
the detector downstream let the undecayed beam particles to pass through. The vacuum ends
at the last station of the spectrometer and downstream the beam passes in vacuum through a
beam pipe. A RICH counter filled with Ne separates $\pi^+, \; \mu^+$ and $e^+$ up to 40 GeV/c.
The time of charged particles is measured both with RICH and with scintillator arrays (CHOD) inherited
from the NA48 experiment, all placed downstream of the RICH. 
The NA48 LKr calorimeter detects forward $\gamma$, and
complements the RICH for particle identifcation. A shashlik small-angle annular calorimeter
(IRC) in front of LKr detects $\gamma$ directed on the inner edges of the LKr hole around the
beam axis. An hadronic calorimeter made by two modules of iron-scintillator sandwiches (MUV1
and MUV2) provides further $\pi^+/\mu^+$ separation and triggering information.
A fast scintillator array (MUV3) identifes and triggers muons with sub-nanosecond time resolution. A shashlik
calorimeter (SAC) placed on the beam axis downstream of a dipole magnet bending off-axis
undecayed beam particles, detects $\gamma$ down to zero angle. Downstream detectors see less
than 1\% of the beam rate. Kaon decays in the vacuum region and particles from upstream beam
activity are the main sources of flux of particles downstream. A multi-level trigger architecture
is used. Timing information from CHOD, RICH and MUV3 and, calorimetric information from
electromagnetic and hadronic calorimeters, are processed by FPGAs' mounted on
TEL62 readout boards\cite{ref:tel}. This infromation produce special data packets used to generate level zero trigger. Software-based information from
KTAG, LAV and magnetic spectrometer provides higher level triggers.

\section{Experiment status}

NA62 collected data in 2014 and 2015. The hardware and readout of the detectors have
been commissioned at 10\% of the nominal beam intensity, the beam line up to nominal intensity.
The Gigatracker ran with a partial hardware configuration. Data samples with beam intensity
varying from some percent of the nominal up to the nominal one, were recorded in 2015. Data
at low intensity were triggered with CHOD only. Calorimetric information have been used to trigger $\pi\nu\bar\nu$-like events at higher intensities. In the following
sections a preliminary analysis of a subset of the low intensity sample is described. The analysis of the full 2015 data set is on-going.

\section{Analysis method}
\label{}

\begin{figure}[h!]
\centering
\includegraphics[width=8cm]{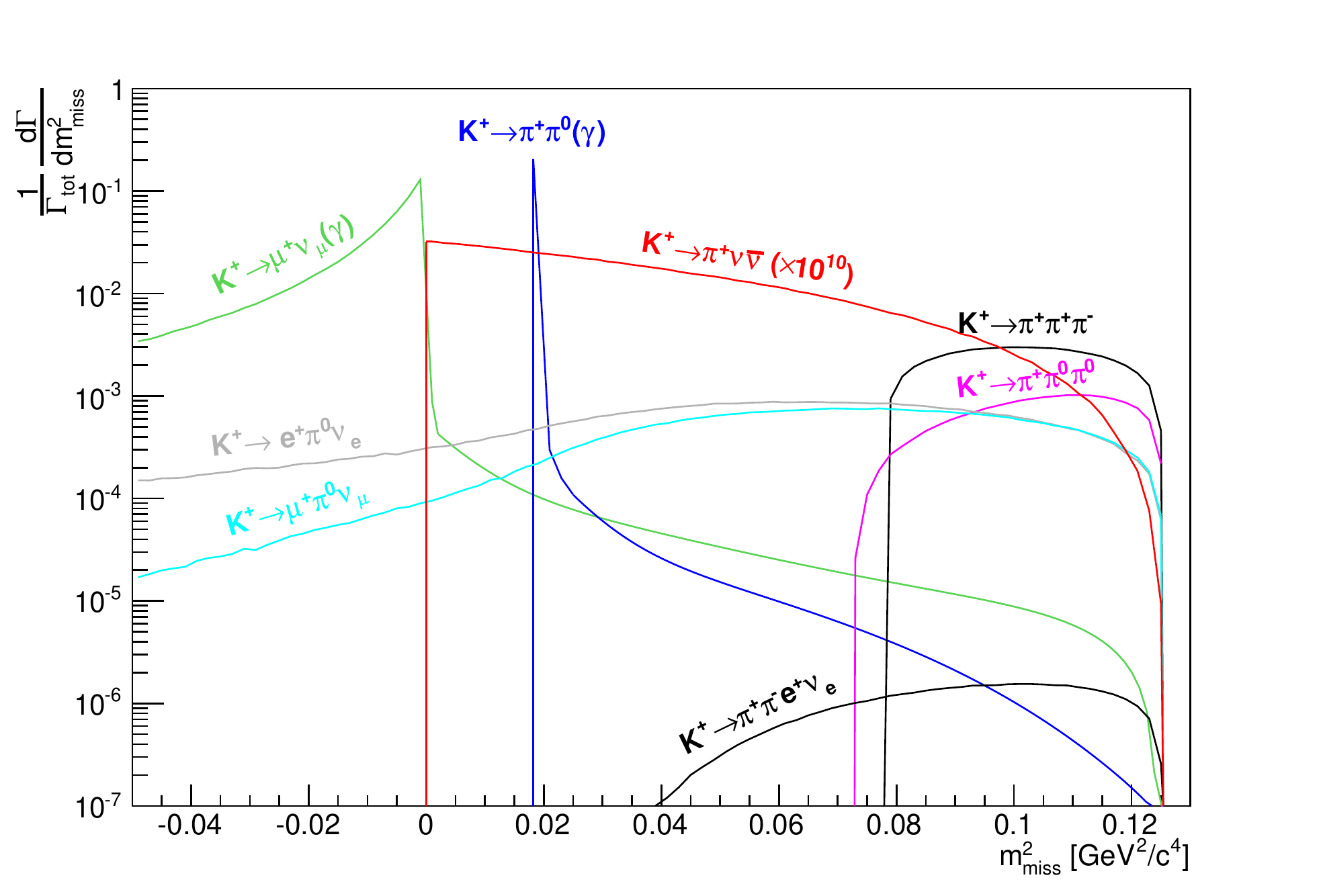}
\caption{Theoretical $m^2_{miss}$ distribution for signal and backgrounds from the main $K^+$
decay modes: the backgrounds are normalized according to their branching ratio; the signal is multiplied by a factor $10^{10}$.}\label{fig:mass}
\end{figure}

\begin{figure}[h!]
\centering
\includegraphics[width=8cm]{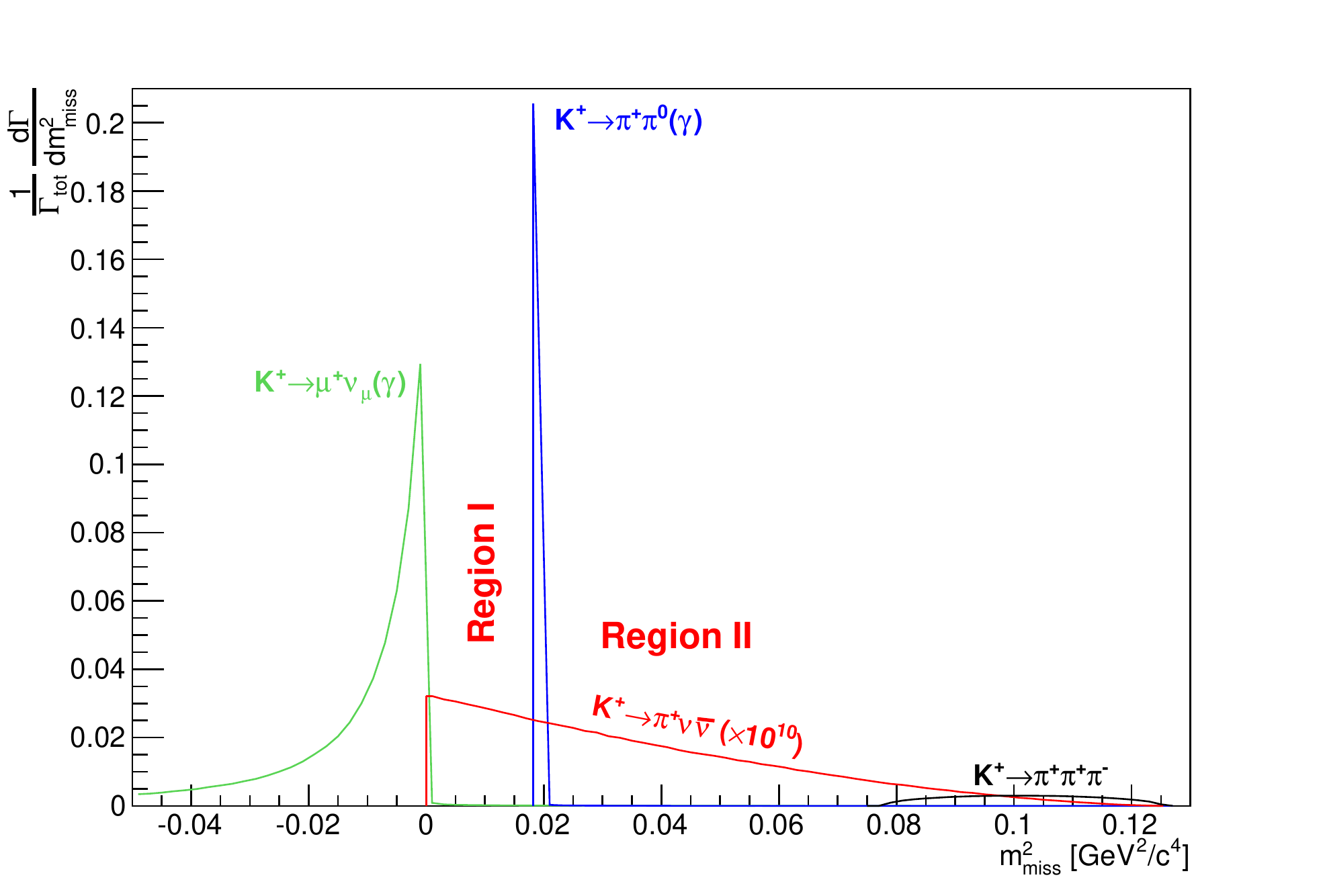}
\caption{theoretical $m^2_{miss}$ distribution for signal and the main background sources after kinematic cuts; the signal is multiplied by a factor $10^{10}$.}\label{fig:mass2}
\end{figure}

The signature of the signal is one track in the Gigatracker compatible with a kaon
and one in the detector downstream compatible with pion. Kaon decays and beam related activity are sources of background.
Let $P_K$ and $P_{\pi^+}$ be the 4-momenta of the kaon and the charged particles produced from 
kaon decay under the $\pi^+$ mass hypothesis, respectively.
The squared missing mass distribution
of the signal, $m^2_{miss} = (P_K - P_{\pi^+})^2$, has a three body decay shape, while more than 90\% of the
charged kaon decays are mostly peaking, as shown in figure \ref{fig:mass}. Signal is looked for in two signal
regions around the $K^+\rightarrow \pi^+\pi^0$ peak. Semileptonic decays, radiative processes, main kaon decay modes via reconstruction tails and beam induced tracks span across these regions. Therefore
kinematic reconstruction, photon rejection, particle identfication, and sub-nanoseconds timing
coincidences between subdetectors must be employed to reduce background.
Figure \ref{fig:mass2} shows the $m^2_{miss}$ distribution for the signal and the three main background sources that survive kinematic cuts. A tight requirement on $P_{\pi^+}$ between 15 and 35 GeV/c boosts the background suppression further, as will be also shown in the next section. 

\section{2015 data quality analysis and prospects for the $BR(K^+\rightarrow \pi^+ \nu \bar \nu)$ measurement}
\label{}

\begin{figure}[h!]
\centering
\includegraphics[width=8cm]{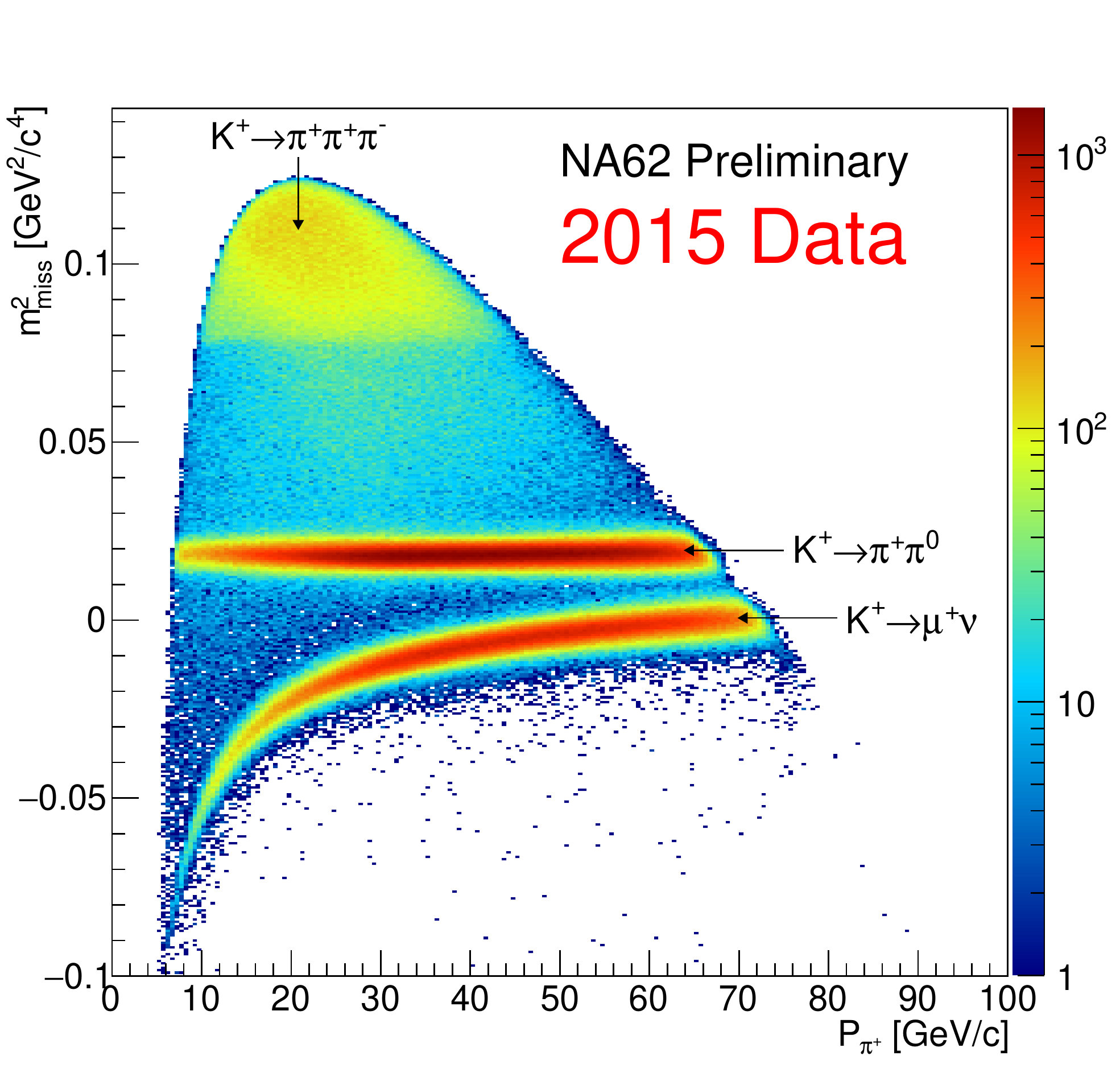}
\caption{$m^2_{miss}$ distribution vs $P_{\pi^+}$ momentum under $\pi^+$ mass hypothesis as a function of the momentum of the track measured in the straw spectrometer after selection for single track from kaon decays.}\label{fig:kaon}
\end{figure}

\begin{figure}[h!]
\centering
\includegraphics[width=8cm]{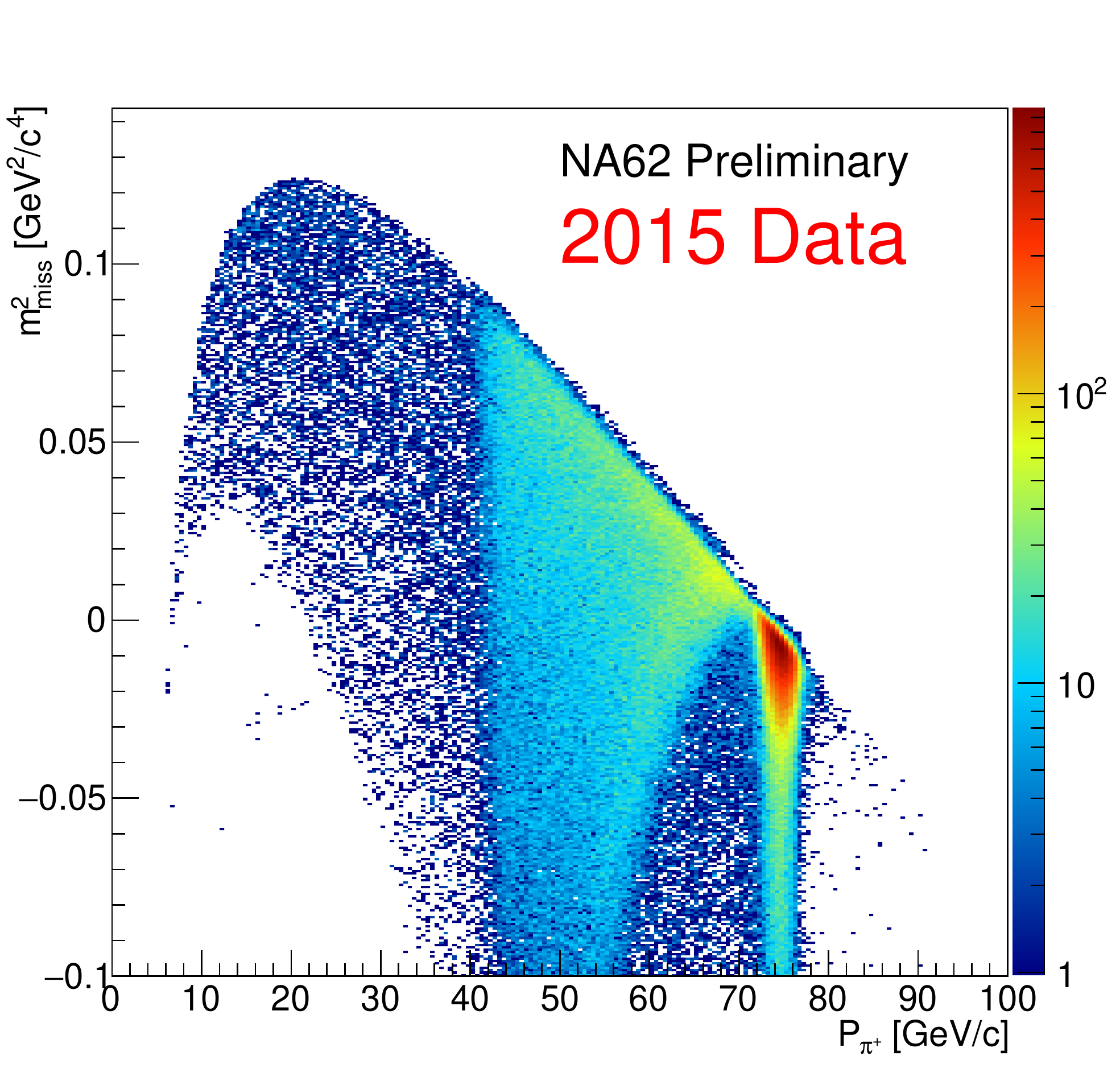}
\caption{$m^2_{miss}$ distribution vs $P_{\pi^+}$ momentum under $\pi^+$ mass hypothesis as a function of the momentum of the track measured in the straw spectrometer asking for single track without a positive kaon tag in time in KTAG.}\label{fig:nokaon}
\end{figure}

The following selection is applied to study the quality of the data for the $K^+\rightarrow \pi^+\nu\bar\nu$ measurement. Tracks reconstructed in the straw spectrometer matching in space energy depositions in
calorimeters and signals in the CHOD are selected. Matched CHOD signals define the time of the
tracks with 200 ps resolution. A track not forming a common vertex within the decay region
with any other in-time track defines a single track event. The last Gigatracker station and the
first plane of the straw spectrometer bound the decay region. A vertex is defined as the average
position of two tracks projected back in the decay region at the closest distance of approch (CDA) less than 1.5 cm. In order to select a single track event originated from kaon decays, a Gigatracker track is required to match the downstream track both in time and space, 
forming a vertex extrapolated into the decay region, 
and to be in-time also with a kaon-like signal in KTAG. Figure \ref{fig:kaon} shows the $m^2_{miss}$ versus the spectrometer track momentum for 2015 data recorded at low intensity. The time resolutions of the KTAG and of a Gigatracker track has been measured to be around 100 and 200 ps, respectively, matching the design values. The KTAG can be used in anti-coincidence with a Gigatracker track to select single track events not related to kaons (Figure \ref{fig:nokaon}).
This technique shows that decays from beam $\pi^+$, elastic scattering of beam
particles in the material along the beam line (KTAG and Gigatracker stations), and inelastic
scatterings in the last Gigatracker station are the main sources of tracks downstream originating
from beam-related activity. The sample of single track events from kaon decays selected above
is used to study kinematic resolution, particle identification, and background rejection.

\begin{figure}[h!]
\centering
\includegraphics[width=8cm]{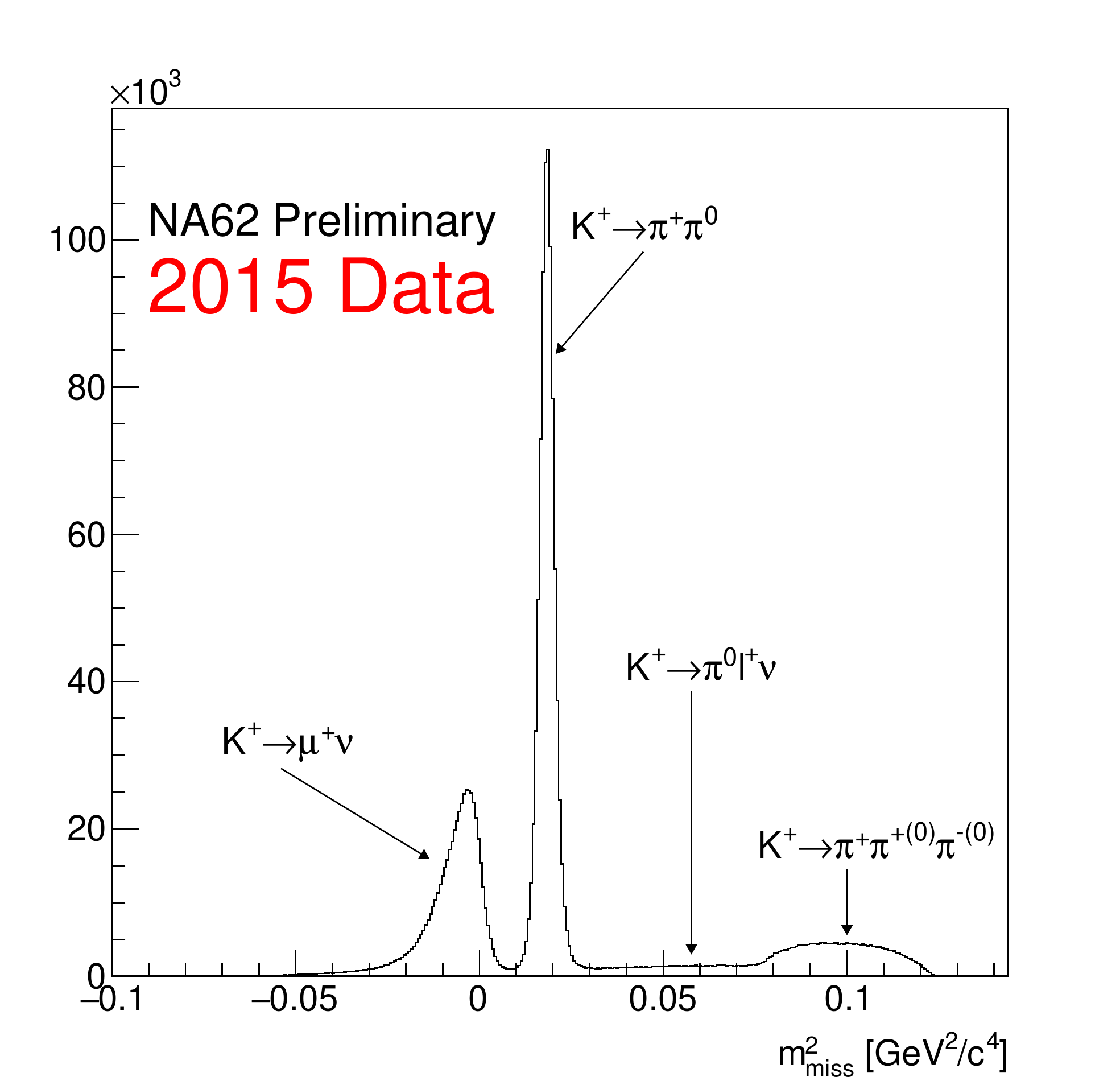}
\caption{$m^2_{miss}$ distribution after the signal event selection. The kaon decays contributing to the distribution are indicated.}\label{fig:final}
\end{figure}

\begin{figure}[h!]
\centering
\includegraphics[width=8.5cm]{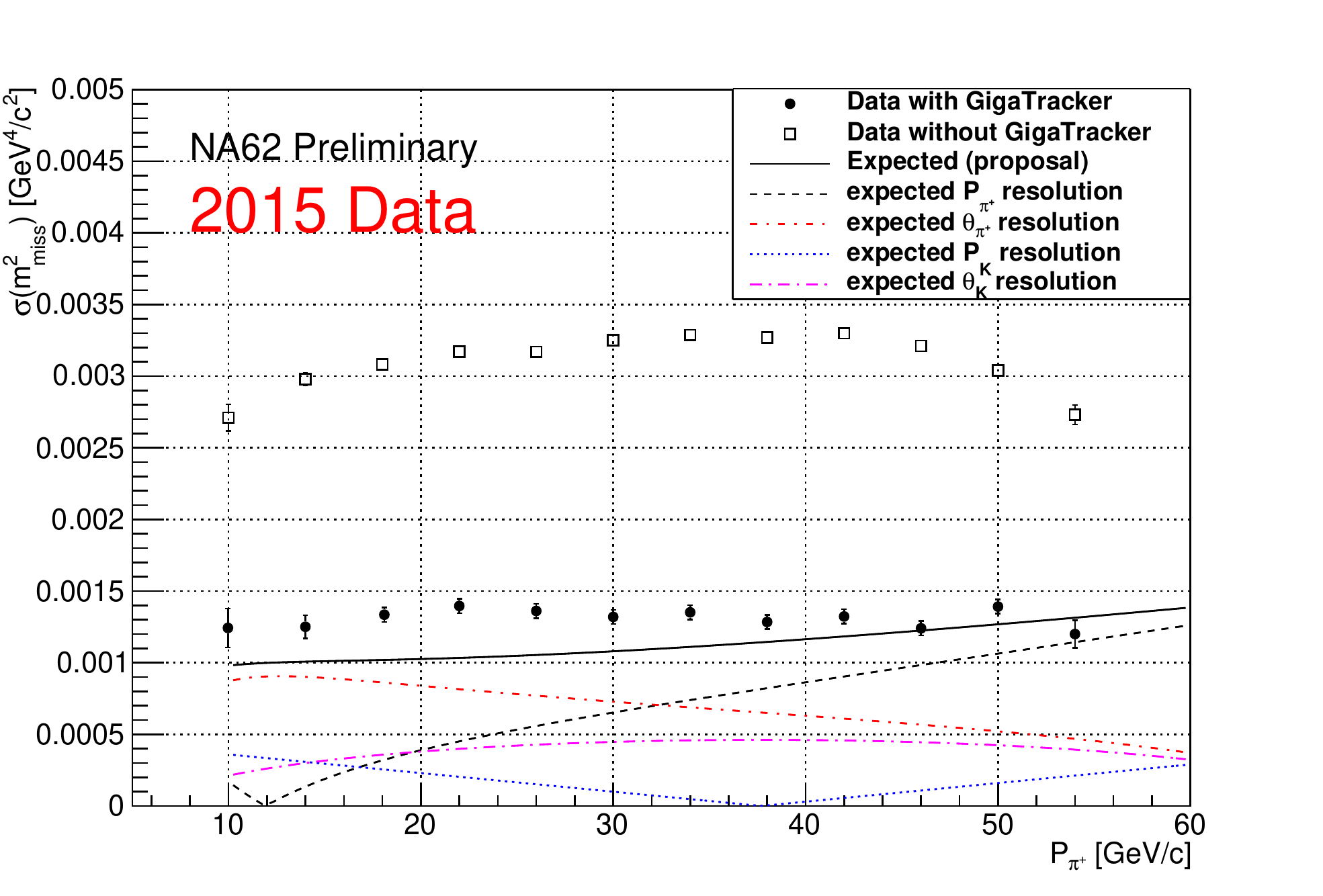}
\caption{$\sigma(m^2_{miss})$ distribution obtained with kaon momentun information from GTK (black points) and with nomimal value (empty squares). Expected resolutions are also shown}\label{fig:sigma}
\end{figure}

The squared missing mass distribution after the analysis kinematic cuts is shown in Figure \ref{fig:final}.
The resolution of the $m^2_{miss}$ measured from the width of the $K^+\rightarrow \pi^+\pi^0$ 
peak is in the range of $1.2 \times 10^{-3}$ GeV$^2/c^4$, close to the $10^{-3}$ GeV$^2/c^4$ design value. The resolution is a factor 3 larger if the nominal kaon momentum is taken, instead of the event by event Gigatracker measured value (Figure \ref{fig:sigma}). The tracking system of NA62 is also designed to provide a rejection factor in the range of $10^4 \div 10^5$ for $K^+\rightarrow \pi^+\pi^0$ and $K^+\rightarrow \mu^+\nu_\mu$ using $m^2_{miss}$ to separate signal from backgrounds, respectively. 
The $K^+\rightarrow \pi^+\pi^0$ kinematic suppression is measured using a subsample of single track events from kaon decays selected by requiring the additional presence of two $\gamma$s compatible with a $\pi^0$ in the LKr calorimeter. This constraint defines a sample of $K^+\rightarrow \pi^+\pi^0$
with negligible background even in the signal $m^2_{miss}$ regions, allowing the study
of the far tails of the $m^2_{miss}$. The measured $K^+\rightarrow \pi^+\pi^0$ suppression factor is of the order of $10^3$. The partial hardware Gigatracker arrangement used in 2015 mainly limits the suppression
because of $m^2_{miss}$ tails due to beam track misreconstruction.

The particle identification of NA62 is designed to separate $\pi^+$ from $\mu^+$ and $e^+$ in order to
guarantee at least 7 order of magnitude suppression of $K^+\rightarrow \mu^+\nu_\mu$in addition to the kinematic rejection. RICH and calorimeters together are employed to this purpose. The pure samples
of $K^+\rightarrow \pi^+\pi^0$ used for kinematic studies and an additional sample of $K^+\rightarrow \mu^+\nu_\mu$ selected among the single track events from kaon decays by asking for the presence of signals in MUV3 are used to study the $\pi^+/\mu^+$ separation in the RICH. About $10^2$ muon suppression for 80\% $\pi^+$ efficiency is measured in a track momentum region between 15 and 35 GeV/c. Above 35 GeV/c the separation degrades quickly as expected from the Cerenkov threshold curves for $\pi^+$ and $\mu^+$ in Neon. As a by-product, the RICH provides an even better separation between $\pi^+$ and $e^+$. The same $\pi^+$ and $\mu^+$ samples allow the calorimetric muon-pion separation to be investigated. Simple cut and count analysis provide a muon suppression factor within $10^4 \div 10^6$ for a 
$\pi^+$ efficiency between 90\% and 50\%. Several analysis techniques are under study to get the optimal separation.

\begin{figure}[h!]
\centering
\includegraphics[width=8.5cm]{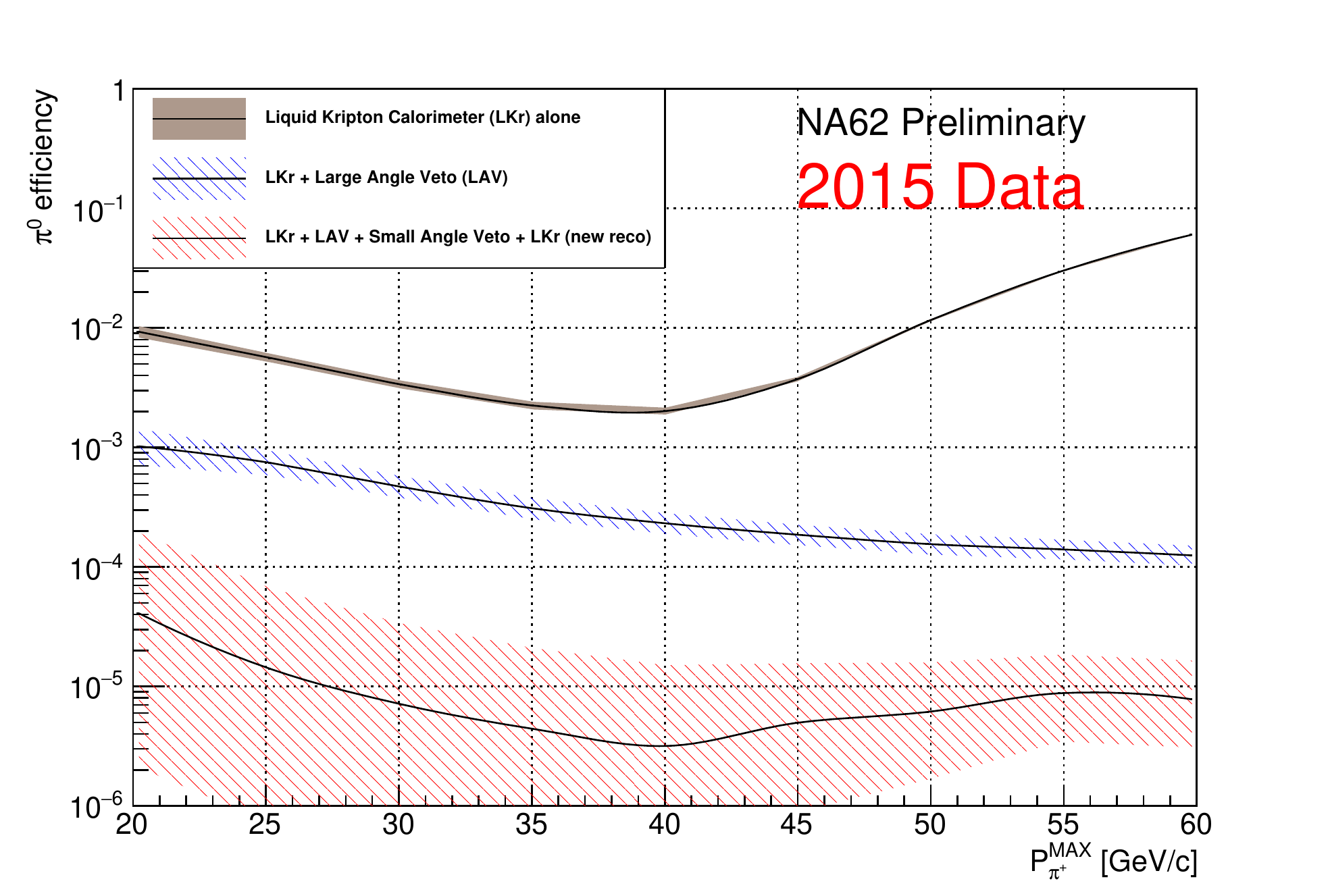}
\caption{$\pi^0$ veto efficiency for different combinations of the calorimeters information.}\label{fig:pi0}
\end{figure}

The layout of NA62 is designed to suppress $K^+\rightarrow \pi^+\pi^0$ of 8 orders of magnitude by detecting at least one photon from $\pi^0$ decay in one of the electromagnetic calorimeters, LAV, LKr and IRC and SAC covering an angular region between $50 \div 8.5$ mrad, $8.5 \div 1$ mrad, $< 1$ mrad, respectively. The $\pi^0$ suppression benefits from the angle-energy correlation between the two $\pi^0$ photons and the cut at 35 maximum $\pi^+$ momentum at analysis level. This allows the suppression factor to be achieved with single photon detection inefficiencies within the reach of the calorimeters,
not below $10^{-5}$ for $\gamma$ above 10 GeV. The suppression of $\pi^0$ from $K^+\rightarrow \pi^+\pi^0$ is measured looking directly at the scaling of the $K^+\rightarrow \pi^+\pi^0$ $m^2_{miss}$ peak after applying conditions for photon rejection on the sample of single track events from kaon decays selected as above. 
The $\pi^0$ veto efficiency measured with the 2015 data, shown in Figure \ref{fig:pi0}, is statistically limited at $10^{-6}$
(90 \% CL) as an upper limit. The signal efficiency is within 90\% as measured on samples of muons from
$K^+\rightarrow \mu^+\nu_\mu$ and events with a single 75 GeV/c  $\pi^+$ in the final state entering in the downstream detector acceptance via beam elastic scattering upstream.

To conclude, the preliminary analysis of low intensity 2015 data shows that NA62 is
approaching the design sensitivity for measuring BR($K^+\rightarrow \pi^+ \nu\bar\nu$).

\section{NA62 physics program beyond $K^+\rightarrow \pi^+ \nu \bar \nu$}
\label{}

The performance of the apparatus allows many physics opportunities beyond the $K^+\rightarrow \pi^+ \nu\bar\nu$ to be addressed. NA62 can study standard $K^+$ decay modes with unprecedented precision,
like the relative measurements of the branching ratios of the more abundant $K^+$ decays or the
measurement of $R_K$. The single event sensitivity offers the possibility to significantly
improve existing limits on lepton flavour and number violating decays like 
$K^+\rightarrow \pi^+\mu^\pm e^\pm$ or $K^+\rightarrow \pi^+l^+l^-$.
Experimentally, $\pi^0$ physics can take advantage
of the performance of the electromagnetic calorimeters, and processes like  $\pi^0 \rightarrow $ invisible,
$\pi^0\rightarrow 3,4 \gamma$ or dark photon
production can be investigated. Thanks to the quality of the kinematic reconstruction, searches for a
heavy neutrino produced in $K^+\rightarrow l^+\nu$  decays can be extend in the heavy neutrino mass region of 300 - 380 MeV/$c^2$ strongly improving the existing limits. The longitudinal scale of the apparatus
and the resolution of the detectors open the possibility to search for long living particles
through their decays, like dark photons decaying in $l^+l^-$ or axion-like particles decaying in $\gamma\gamma$ pairs, whether produced at the target or in beam dump configurations.





\nocite{*}
\bibliographystyle{elsarticle-num}
\bibliography{jos}



\end{document}